\documentclass[11pt]{article}
\usepackage{amsmath,amsfonts, amsthm} 
\usepackage{multirow}

\newcommand{\me}{\mathrm{e}}
\newcommand{\mi}{\mathrm{i}}
\newcommand{\dif}{\mathrm{d}}

\newcommand{\mathematica}{\emph{Mathematica}}

\renewcommand{\Im}{\operatorname{Im}}

\newtheorem{theorem}{Theorem}
\newtheorem{lemma}[theorem]{Lemma}
\newtheorem{proposition}[theorem]{Proposition}
\newtheorem{corollary}[theorem]{Corollary}
\theoremstyle{definition}
\newtheorem{defi}[theorem]{Definition}
\theoremstyle{remark}
\newtheorem{remark}{Remark}

\newcommand{\Mvariable}[1]{#1}
\newcommand{\todo}[1]{\vspace{5 mm}\par \noindent
	\marginpar{\textsc{ToDo}}
	\framebox{\begin{minipage}[c]{0.92 \textwidth}
	\tt #1 \end{minipage}}\vspace{5 mm}\par}
\newcommand{\note}[1]{\vspace{5 mm}\par \noindent
	\marginpar{\textsc{Note}}
	\framebox{\begin{minipage}[c]{0.92 \textwidth}
	\tt #1 \end{minipage}}\vspace{5 mm}\par}
\newlength{\leftlablen}
\newcommand{\leftlabel}[1]{%
	\settowidth{\leftlablen}{\framebox{#1}}
	\addtolength{\leftlablen}{0.5in}
	\hspace{-\leftlablen} \framebox{#1} \\[-5.58ex]}
\renewcommand{\todo}[1]{}
\renewcommand{\note}[1]{}
\renewcommand{\leftlabel}[1]{}

\title{Collisions of Four Point Vortices in the Plane}
\author{Antonio Hern\'andez-Gardu\~no \and Ernesto A. Lacomba}
\date{This version September 6, 2006}

\begin{document}

\maketitle

\begin{abstract}
	This paper addresses the question of existence of (not necessarily self-similar) solutions to the $4$-vortex problem that lead to total or partial collision.  We begin by showing that energy considerations alone imply that, for the general $N$-vortex problem, the virial being zero is a necessary condition for a solution to both evolve towards total collision and satisfy certain regularity condition.  For evolutions assumed to be bounded, a classification for asymptotic partial collision configurations is offered.  This classification depends on inertia and vorticity considerations.  For non-necessarily bounded evolutions, we discuss the relationship between partial and (non-regular) total collisions when the virial is not zero and a generic condition on the vorticities holds.  Finally, we give a canonical transformation that, for a class of $4$-vortex systems satisfying a restriction on the vorticities, allows to formally apply the averaging principle in order reduce the dynamics when two of the vorticities are near a binary collision.  The reduced system is a one-degree of freedom hamiltonian system.
\end{abstract}

\section{Introduction} 

The study of the dynamics of point-vortices on two dimensional manifolds, for an ideal incompressible fluid, has become a very active field of research in recent years.  For background on the subject we refer the reader to the comprehensive book of P. Newton (2000) \cite{Newton2001}.  The case of vortices on the plane has a long history going back to W. Gr\"obli's 19th century work (cf. \cite{ArefRottThomann1992}), who studied the analytical solutions of the three vortex problem for certain special cases.  Further developments on the three-vortex problem were carried out in \cite{Aref1979, Chapman1978, Novikov1975, NovikovSedov1979, Synge1949}.  In \cite{HernandezLacomba2005} (see also \cite{Aref1979}) we prove that total collision for three point vortices on the plane is only possible when the virial (defined below) is zero, in which case the vortices evolve in a self-similar fashion.  In op. cit. we also prove that partial (i.e. binary) collision can not happen with only three vortices on the plane.

In this paper we are interested in the problem of existence of total or partial collapse for a system of four point-vortices on the plane.  
The motion of four vortices in an arbitrary configuration, but all with equal vorticity, has been studied in \cite{ArefPomphrey1982} (see also \cite{ArefPomphrey1980}), both analytically and numerically, where it is shown that the four-vortex problem is not integrable.  
In that paper a canonical transformation reducing the system from $N$ equal vortices to a system with $ N - 2 $ degrees of freedom is found.  A similar transformation is shown to exist in \cite{EckhardtAref1988} for the dynamics of two vortex dipoles where the total vorticity of each dipole is zero and an analytical study of collision trajectories and a numerical study of the scattering in both the regular and the chaotic regime is offered.

To our knowledge, all collisions solutions to the $N$-vortex problem studied in the literature involve \emph{self-similar} evolutions.  It is known (\cite{NovikovSedov1979}) that these can only occur when the \emph{virial} is zero.  So a first natural question is whether a null virial is a necessary condition for total collision.  We answer this question affirmatively for a class of collisions that are \emph{regular} in a precise sense.  To illustrate this result, we apply it to the problem of four vortices in parallelogram configuration, first studied by Novikov and Sedov in op. cit.  After defining carefully the different senses in which a system can develop partial collisions, we determine necessary conditions for partial collapse and necessary asymptotic configurations under the assumption of a bounded evolution.  Then, for non-necessarily bounded evolutions we study the relationship between partial and (non-regular) total collision, classifying this relationship in terms of the sign of the vorticity.  In this context a non-zero virial and a generic condition on the vorticities is assumed.

Some of the results described in the previous paragraph motivate the study of the evolution of a $4$-point vortex system in a configuration where two of the vortices (say $ \Gamma _1 $ and $ \Gamma _2 $) are close to binary collapse.  This is a complicated problem but we believe that averaging methods can give useful information.  A first step in this direction is achieved in the last part of this paper, where we describe a canonical transformation that, in the class of systems satisfying the condition $ \Gamma _1 + \Gamma _2 = \Gamma _3 = \Gamma _4 $, allows to formally apply the averaging principle in order to reduce the $4$-vortex system to a one-degree of freedom hamiltonian system.

\pagebreak
\subsection{Preliminaries}

The system of $N$ point-vortices on the plane can be described as follows.  Let $ z _\alpha = x _\alpha + \mi \, y _\alpha \in \mathbb{C} $ be the coordinates of the $ \alpha $-th point vortex with vorticity $ \Gamma _\alpha $, $ \alpha = 1, \ldots , N $.  Thus the configuration space is $ \mathbb{C} ^n \cong \mathbb{R} ^{ 2n} $.  The equations of motion are given by
\begin{equation}\label{dynecs0}
	\dot{ z} _\alpha = \frac{ \mi }{ 2 \pi} \sum _{ \beta \neq \alpha}^N \Gamma _\beta \frac{ z _\alpha - z _\beta }{ |z _\alpha - z _\beta | ^2 }
\end{equation}
This system is Hamiltonian, with the energy (Hamiltonian) function given by 
\begin{equation}\label{energy_1}
  H = - \frac{ 1 }{ 2 \pi } \sum _{ \alpha < \beta} \Gamma _\alpha \Gamma _\beta \ln l _{ \alpha \beta} \;, 
\end{equation} 
where $ l _{ \alpha \beta} := | z _\alpha - z _\beta | $, and symplectic form given by 
\begin{equation} \label{symplectic_form}
  \Omega (z, w) = - \operatorname{ Im} \sum _{ \alpha = 1} ^n \Gamma _\alpha z _\alpha \bar{ w} _\alpha \;,
\end{equation}
where the $ \Gamma _\alpha \in \mathbb{R} $, $ \Gamma _\alpha \neq 0 $.  Hence $H$ is a conserved quantity, but also some other conserved quantities arise.  From \eqref{dynecs0} it follows that
\[
  Z = \sum _\alpha \Gamma _\alpha z _\alpha 
\] 
and
\[
  I = \sum _\alpha \Gamma \rho _\alpha ^2 \;, \quad \rho _\alpha = | z _\alpha |
\] 
are conserved as well.  Here $ \Gamma = \sum _\alpha \Gamma _\alpha $.

If $ \Gamma \neq 0 $ we call $ Z/\Gamma $ the barycenter of the system.  A calculation shows that
\begin{equation} \label{Meq}
  M := \sum _{ \alpha < \beta} \Gamma _\alpha \Gamma _\beta \, l _{ \alpha \beta} ^2 = \Gamma I - |Z| ^2 \;. 
\end{equation}
Thus, conservation of $Z$ and $I$ imply conservation of $M$.  Finally, from \eqref{dynecs0} it also follows that the angular momentum or \textbf{virial}
\[
  V = \sum _\alpha \Gamma _\alpha \, z _\alpha \times \dot{z}_\alpha =\sum _{\alpha<\beta} \Gamma _\alpha \Gamma _\beta
\] 
is an invariant of the system, i.e., a constant depending only on the vorticities.
Without loss of generality we will assume that $ Z \equiv 0 $ when $\Gamma\neq 0$, and hence $ M = \Gamma I $.  

Notice that all initial configurations that end in total collision must satisfy $ M = I = 0 $ (even when $\Gamma = 0$). 

We say that the dynamical evolution of $N$ vortices on the plane is a \textbf{self-similar} trajectory if it can be expressed as $ z _i (t) = \zeta (t) z _i (0) $, $ i = 1, \ldots N $, for some $ \zeta : \mathbb{R} \longrightarrow \mathbb{C} $ (hence $ Z = 0 $).




\subsection{Total and partial collapse:  definitions and generalities}
\label{ssec:defs_gens_collapse}

\note{A:  Below we define the notion of collision.  Should be put in context.}

\leftlabel{s-DC}

In order to be able to state the results in later sections, we must be precise on what we mean by total or partial collision in the $N$-vortex problem.  (We remark that the following definitions are independent of the specific dynamic structure of point-vortex systems inherited from Euler's equation.)

\begin{defi}
	We say that a system of $N$ point-vortices (with positions $ z _i \in \mathbb{C} $, $ i = 1, \ldots, N $) evolves towards an \textbf{$n$-collision} (\textbf{binary, ternary}, etc.~-collision if $ n = 2, 3 $, etc.), or simply a \textbf{collision}, in time $ T _\ast \le \infty $ if there are indices $ I = \{ i _1, \ldots, i _n \} $, $ n \le N $, such that for every $ i, j \in I $ we have that $  | z _i - z _j | \rightarrow 0 $ as $ t \rightarrow T _\ast $. We say that the system evolves towards a \textbf{sequential $n$-collision} (binary, ternary, etc.), or simply a \textbf{sequential collision}, if there is a monotonically increasing sequence of times  $ \{ t _k \} $ converging to $ T _\ast \le \infty $ such that $ | z _i  (t _k) - z _j (t _k) | \rightarrow 0 $ as $ k \rightarrow \infty $ for all $ i,j \in I $.
\end{defi}

It has been shown in \cite{HernandezLacomba2005} that in the three-vortex problem there are no binary collisions.  However, it makes sense to consider the existence of partial collisions in the $N$-vortex problem.  To this end we state with the following

\begin{defi}
	If in the previous definition 
	$ n = N $ we say that a system evolves towards a \textbf{total collision}.  Assuming that the vortices labelled $ i _1, \ldots, i _n $, $ n < N $, evolve towards a total collision, we call it a \textbf{proper $n$-collision} (proper- binary, ternary, etc.), or simply a \textbf{proper partial collision}, if there are no other vortices labelled $ i _{n+1}, \ldots, i _{ n+k} $, $ k \le N-n $, such that the whole set of vortices $ i _1, \ldots, i _{ n+k} $ evolve towards a total collision.  A \textbf{proper sequential $n$-collision} is defined analogously.
\end{defi}

It also makes sense to talk about partial collisions in the context of similarity classes of $N$ point-vortices on the plane.  So, for example, it makes sense to speak of a binary collision within a process of total collision or blow-up.  To define these types of collisions we first introduce a notation for the rescaling of the (squared) distance between the point-vortices:  

If $ (z _1 (t), \ldots, z _N (t)) $ is a trajectory in $ \mathbb{C} ^N $ describing the evolution of a system of $N$ point-vortices, let $ b _{ ij} := | z _i - z _j | ^2 $ and let the parameters $ \{ \rho, \beta _{ ij} \} $ be defined by the conditions $ b _{ ij} = \rho \, \beta _{ ij} $, $ \rho > 0 $, $ \sum _{ i<j} \beta _{ ij} = 1 $.  With this notation we state the following

\begin{defi}
	We say that a system of $N$ point-vortices evolves towards a \textbf{relative $n$-collision} (relative- binary, ternary, etc.), or simply a \textbf{relative partial collision}, in time $ T _\ast \le \infty $ if there are indices $ I = \{ i _1, \ldots, i _n \} $, $ n < N $, such that for every $ i, j \in I $ we have that $ \beta _{ ij} \rightarrow 0 $ as $ t \rightarrow T _\ast $.  A \textbf{relative sequential $n$-collision} is defined analogously.
\end{defi}

The term \textbf{proper relative (sequential) $n$-collision} is defined in an obvious way.

\todo{A: Incorporate explicitly the definition of proper relative n-collision in the previous definition.}

\begin{remark}
	It is clear that $ \{ $proper $n$-collisions$ \} \subset \{ $proper relative $n$-collisions$\} \subset \{ $relative $n$-collisions$ \}  \subset \{ n$-collisions$ \} $.
\end{remark}

In this paper we consider the question of the existence of evolutions of the four vortex problem that lead to total collision when the virial is not zero.  From the energy relation \eqref{energy_1} it follows that if total collision is to occur then the vorticities can not all have the same sign.  The two possible cases are: a) three vorticities with the same sign and one with opposite and b) two positive and two negative vorticities.  


Let $ b _{ ij} := | z _i - z _j | ^2 $.  Then a quadrilateral configuration is represented by a point $ (b _{ 12}, b _{ 13}, b _{ 14}, b _{ 23}, b _{ 24}, b _{ 34}) \in \mathbb{R} ^6 $, subject to the constraint that the Cayley-Menger determinant $ \det (V _{ ij}) $ vanishes, where $ (V _{ ij}) $ is the $ 5 \times 5 $ matrix given by 
\[
  V _{ ij} = \left\{ \begin{array}{rl}
  	| \delta_{ 1i} - \delta_{ 1j} | & \text{if $ i = 1 $ or $ j = 1 $} \,, \\
	b _{ (i-1)(j-1) } & \text{if $ i > 1 $ and $ j > 1 $} \,,
  \end{array} \right.
\] 
where $ \delta _{ ij} $ is the Kronecker delta, representing $ 288 $ times the square of the volume of the tetrahedron with sides of length $ |z _i - z _j|  $ (cf. \cite[p.  256]{Uspensky1948}).

Through out this paper we use the rescaling coordinates $ \{  \rho,\beta _{ij} \} $ defined by $ b _{ ij} = \rho\, \beta _{ ij} $, subject to the constraint $ \sum \beta _{ ij} = 1 $.  Notice that to each point $ (\beta _{ ij}) \in \mathbb{R} ^6 $ satisfying both $ \sum \beta _{ ij} = 1 $ and $ \det (V _{ ij}) = 0 $ we can associate a point in the space of \emph{quadrilateral shapes}, understood as the set of equivalence classes in the set of all quadrilaterals, with equivalence given by similarity, and that this correspondence is a bijection.

Exponentiating both sides of \eqref{energy_1} we arrive at
\begin{equation} \label{energy_2}
	\frac{ h }{ \rho ^{ V}} = 
		\beta _{ 12} ^{ \Gamma _1 \Gamma _2}\, 
		\beta _{ 13} ^{ \Gamma _1 \Gamma _3}\, 
		\beta _{ 23} ^{ \Gamma _2 \Gamma _3}\, 
		\beta _{ 14} ^{ \Gamma _1 \Gamma _4}\, 
		\beta _{ 24} ^{ \Gamma _2 \Gamma _4}\, 
		\beta _{ 34} ^{ \Gamma _3 \Gamma _4} \,. 
\end{equation}

\section{Dynamics and total collapse for $N$ point-vortices}

\subsection{Dynamics in the space of squared distances}
\label{ssec:squared_distances_dynamics}

The discussion of the collapse of $N$ point-vortices on the plane is best dealt with by describing the dynamics in the space of squared-distances between the vortices.  This space is given by the set $\mathcal{Q}$ of points $ (b _{ 12}, \ldots, b _{ (N-1)N}) \in \mathbb{R} ^{ N \choose 2 } $ such that $ b _{ ij} := |z _i - z_j| ^2 $, where  $ (z _1, \ldots, z _N) \in \mathbb{C} ^N $.  Thus, $\mathcal{Q}$ is the set of vectors in $ \mathbb{R} ^{ N \choose 2 } $ with non-negative entries such that, for every triad $ (i,j,k) $ of vortex labels, $ (b _{ ij}, b _{ jk}, b _{ ki}) $ satisfies the triangle inequality
\[
	b _{ij} ^2 + b _{jk} ^2 + b _{ki} ^2 \le  2(b _{ij} b _{jk} + b _{ij} b _{ki} + b _{jk} b _{ki}) \;.
\]

Let $\mathcal{C}$ be the collision subset of $\mathcal{Q}$.  That is to say, $\mathcal{C}$ is the set of points $ (b _{ 12}, \ldots, b _{ (N-1)N}) \in \mathcal{Q} $ such that $ b _{ ij} = 0 $ for at least one pair of vortex-labels $ (i,j) $.

The dynamics in $\mathcal{Q}$ is given by the following

\begin{proposition}\label{prop:dynecs}
	For a system of point-vortices $ \{ \Gamma _1, \ldots, \Gamma _N \} $ the equations of movement are given by
	\begin{equation}\label{sqdist_dynecs}
		\dot{ b} _{ ij} = \frac1{ 2 \pi} \sum _{ k \neq i,j} \Gamma _k \, A _{ ikj} \left( \frac1{b _{ ik}} - \frac1{b _{ kj}} \right) \,, \quad i < j \,.
	\end{equation} 
	Here $ b _{ ij} = |z _i - z _j |^2 $, $ z _i \in \mathbb{C} $ denotes de position of the $i$-th vortex and $ A _{ ikj} $ is the oriented area of the triangle $ \triangle (z _i, z _k, z _j) $.
\end{proposition}

For the proof of this proposition we need the following

\begin{lemma}\label{lem:identity1}
	The identity
	\[
		\Im \left( \frac{ z _1 }{ z _2 } + \frac{ z _2 }{ z _1 } \right) \equiv \Im \left( \bar{ z} _1 z _2 \right) \left( \frac1{ | z _1 | ^2} - \frac1{ | z _2 | ^2 } \right)
	\] 
	holds for all $ z _1, z _2 \in \mathbb{C} \setminus \{0\} $.
\end{lemma}

\begin{proof}
	Express the $ z _i $ in polar coordinates.
\end{proof}

\begin{proof}[Proof of proposition \ref{prop:dynecs}]
	Let $ w _{ ij} = z _i - z _j $ for $ i \neq j $.  Then 
	\[
	  \dot{b} _{ ij} = \frac{ \dif }{ \dif t} \left( w _{ ij} \bar{ w} _{ ij} \right) = \dot{ w} _{ ij} \bar{ w} _{ ij} + w _{ ij} \dot{ \bar{ w}} _{ ij}  \,.
	\]
	From \eqref{dynecs0},
	\[
		\dot{ w} _{ ij} = \frac{ \mi }{ 2 \pi} \left[ \sum _{ k \neq i,j} \frac{ \Gamma _k }{ \bar{w} _{ ik}} + \frac{ \Gamma _j }{ \bar{w} _{ ij}} - \sum _{ k \neq i,j} \frac{ \Gamma _k }{ \bar{w} _{ jk}} - \frac{ \Gamma _i }{ \bar{w} _{ ji}} \right] \,.  
	\]
	Hence,
	\begin{align*}
		\dot{w} _{ ij} \bar{w} _{ ij} &= \frac{ \mi }{ 2 \pi} \left[ \sum _{  k \neq i,j} \Gamma _k \left( \frac{ \bar{w} _{ ij} }{ \bar{w} _{ ik}} + \frac{ \bar{w} _{ ij} }{ \bar{w} _{ kj}} \right) + \Gamma _i + \Gamma _j \right] \\
		\intertext{and}
		w _{ ij} \dot{ \bar{w}} _{ ij} &= \frac{ -\mi }{ 2 \pi} \left[ \sum _{  k \neq i,j} \Gamma _k \left( \frac{ w _{ ij} }{ w _{ ik}} + \frac{ w _{ ij} }{ w _{ kj}} \right) + \Gamma _i + \Gamma _j 
		\right] \,.
	\end{align*}
	Summing both equations, while taking into account that
	\[
		\frac{ w _{ ij} }{ w _{ ik}} + \frac{ w _{ ij} }{ w _{ kj}} = 2 + \frac{ w _{ kj} }{ w _{ ik}} + \frac{ w _{ ik} }{ w _{ kj}} \,, 
	\]
	we obtain
	\[
		\dot{b} _{ ij} = \frac{ - \mi ^2 }{ \pi} \sum _{ k \neq i,j} \Gamma _k \Im \left( \frac{ w _{ kj } }{ w _{ ik}} + \frac{  w _{ ik} }{ w _{ kj}} \right) \,. 
	\]
	Applying lemma \ref{lem:identity1} and noting that $ A _{ ikj} = \Im \left( \bar{w} _{ ik} w _{ kj} \right) /2 $ we get the desired result.
\end{proof}

\todo{A: Say some words about the fact that proposition \ref{prop:dynecs} generalizes an analogous proposition in our previous paper and that it is related to an analogous proposition in a paper by P. Newton in the context of vortices on the sphere.}

\subsection{A criterion for total collapse of $N$ point-vortices}
\label{sec:regular_N_vortices_collapse}
\leftlabel{s-RCV0}

We have seen that a condition for total collapse in the $N$ point-vortex problem is that $ M = 0 $.  In this subsection we will show  that, under the assumption that the evolution of the system satisfies certain regularity condition, another necessary condition for total collapse is that $ V = 0 $.  As we will see, the proof of this fact is based solely on energy considerations.

\todo{A:  Discuss whether it is known that, as in the case of three vortices, $ V = 0 $ implies the existence of total collision in the $N$-vortex problem.  Discuss that $ V = 0 $ implies that the evolutions leading to total collapse are self-similar (this comes in Novikov \& Sedov, sect. 2).}

In what follows we will use the following standard notation.  If $ v \in \mathbb{R} ^n \setminus \{ \mathbf{0} \} $ then $ \hat{ v} := v/\|v\| $.  If $ f(t) $ is differentiable then $ \dot{ f} := d f/dt $.

\begin{lemma} \label{lem:regular_curve}
	Let $ \gamma : (a,b) \longrightarrow \mathbb{R} ^n $ be a regular curve not intersecting the origin, $ b \le \infty $, such that $ \lim _{t \rightarrow b} \gamma (t) = \mathbf{0} $.  Then the following are equivalent:
	\renewcommand{\labelenumi}{\roman{enumi}.}
	\begin{enumerate}
		\item The limit of $ \hat{\gamma} (t) $ as $ t \rightarrow b $ exists.
		\item The limit of $ \hat{ \dot{\gamma} } (t) $ as $ t \rightarrow b $ exists.
	\end{enumerate}
	Moreover, if these limits exist then one is the negative of the other.
\end{lemma}

\begin{proof}
	Suppose that \emph{i.} holds.  By L'H\^opital's rule,
	\[
		\lim _{ t \rightarrow b} \hat{\gamma} (t) = \lim _{ t \rightarrow b} \frac{ \dot{\gamma} (t) }{ \frac{d}{dt} \| \gamma (t) \| } = \lim _{ t \rightarrow b} \frac{ \hat{ \dot{\gamma}} (t) }{ \hat{\gamma} (t) \cdot \hat{ \dot{\gamma}}(t) } \,. 
	\]
	The last equality follows from the identity $ \| \gamma \| \frac{d}{dt} \| \gamma \| = \gamma \cdot \dot{\gamma} $, easily obtained by time-differentiating $ \| \gamma \| ^2 $.  Since the limit in the left-hand side exists then
	\[
		1 = \lim _{ t \rightarrow b} \| \hat{\gamma} (t) \| = \lim _{ t \rightarrow b} \frac{ 1 }{ \| \hat{\gamma} (t) \cdot \hat{ \dot{\gamma}} (t) \| } \,. 
	\] 
	Hence $ \lim _{ t \rightarrow b} \hat{\gamma} (t) \cdot \hat{ \dot{\gamma}} (t) = \pm 1 $ and it follows that \emph{ii.} holds.  
	
	Suppose now that \emph{ii.} holds.  Let $ t: (a, c) \rightarrow (a,b) $ be a reparametrization such that $ \bar{\gamma} (\tau) = \gamma(t(\tau)) $ is a curve parametrized by arc length.  If $ c = \infty $ then the fact that $ \lim _{ \tau \rightarrow \infty} \bar{\gamma}\,' (\tau) $ exists, with $ {}' = d/d \tau $, implies that $ \| \bar{\gamma} (\tau) \| $ is unbounded as $ \tau \rightarrow \infty $, contradicting the hypothesis that $ \lim _{ \tau \rightarrow \infty} \bar{\gamma} (\tau) = \mathbf{0} $.  Thus $ c < \infty $ and there exists curve parametrized by arc length $ \bar{\bar \gamma} : (a, c + \epsilon ) \rightarrow \mathbb{R} ^n $, $ \epsilon > 0 $, such that $ \bar{ \bar \gamma}(0) = \mathbf{0} $ and $ \bar{\gamma} = \bar{ \bar \gamma} | (a, c) $.  Then $ \bar{\bar \gamma} (\tau) = - \, \bar{\bar \gamma}\,' (c) (c-\tau) + \rho (\tau) $, where $ \rho (\tau) = o (c-\tau) $.  Hence
	\[
		\lim _{ t \rightarrow b} \hat{\gamma} (t) = \lim _{ \tau \rightarrow c ^-} \widehat{ \bar{\bar \gamma}} (\tau ) = \lim _{ \tau \rightarrow c ^-} \frac{ - \, \bar{\bar \gamma}\,' (c ) + \frac{ \rho (\tau ) }{ c-\tau } }{ \| - \bar{\bar \gamma}\,' (c ) + \frac{ \rho (\tau ) }{ c-\tau } \| } = - \, \widehat{ \bar{\bar \gamma}}\, ' (c) \,, 
	\] 
showing that \emph{i.} holds.  Finally, it is clear that $ \widehat{ \bar{\bar \gamma}}\, ' (c) = \lim _{ t \rightarrow b} \hat{\gamma} (t) $.  Hence the limit in \emph{i.} and minus the limit in \emph{ii.} are equal.
\end{proof}

\begin{defi} \label{def:orig_reg_way}
	Let $ \gamma : (a,b) \rightarrow \mathbb{R} ^n $ be a regular curve not intersecting the origin, $ b \le \infty $.  We say that $ \gamma (t) $ \textbf{approaches the origin in a regular way} as $ t \rightarrow b $ if $ \lim _{ t \rightarrow b} \gamma (t) = \mathbf{0} $ and $ \gamma $ satisfies any (and hence both) of the properties \emph{i.} and \emph{ii.} in lemma \ref{lem:regular_curve}.
\end{defi}

\begin{remark}
	An analogous lemma and definition can be stated for a regular curve $ \gamma : (a, b) \rightarrow \mathbb{R} ^n $ when $ \gamma (t) \rightarrow \mathbf{0} $ as $ t \rightarrow a $.  It is easy to see that the only difference in this case will be that the limits in \emph{i.} and \emph{ii.} of lemma \ref{lem:regular_curve} will be the same.
\end{remark}

We now state the main result of this subsection:

\begin{proposition} \label{prop:reg_collapse_virial}
	Let $ \mathbf{b} (t) = (b _{ 12} (t), \ldots, b _{(N-1)N (t)}) $ be the curve in $ \mathbb{R} ^{ N\choose2} $ that is the solution to the equations of motion \eqref{sqdist_dynecs} for a given initial condition $ \mathbf{b} (0) \in \mathcal{Q} \setminus \mathcal{C} $.  Suppose that there is a time $ T _\ast \le \infty $ such that the curve $ \mathbf{b} (t) $ approaches the origin in a regular way as $ t \rightarrow T _\ast $.  Then $ V = 0 $.
\end{proposition}

\begin{proof}%
The set $ \mathcal{Q} \setminus \mathcal{C} $ is contained in $Q := $~the set of vectors in $ \mathbb{R} ^{ N\choose2} $ with positive entries.  From \eqref{energy_1} we have that, on $Q$, the surfaces of constant energy are given by $ \mathcal{E} _h = F ^{-1} (h) $, where $ h = \exp (-4 \pi H) $ and
\[
	F : Q \longrightarrow \mathbb{R} :
	F(b _{ 12}, \ldots, b _{ (N-1)N} ) := \prod _{ i<j} b _{ ij}^{ \Gamma _i \Gamma _j} \,. 
\]
Let $ \mathbf{b} = ( b _{ 12}, \ldots, b _{ (N-1)N} ) \in Q $, $ \delta \mathbf{b} = ( \delta b _{ l2}, \ldots, \delta b _{ (N-1)N} ) \in T _{ \mathbf{b}} Q $.  Since 
\begin{equation} \label{dif_FF}
	d F _{ \mathbf{b} } \cdot \delta \mathbf{b} = F ( \mathbf{b}) \left( \Gamma _1 \Gamma _2 \frac{ \delta b _{ 12} }{ b _{ 12} } + \cdots + \Gamma _{ (N-1)} \Gamma _N \frac{ \delta b _{ (N-1)N} }{ b _{ (N-1)N} } \right) 
\end{equation}
then $F$ is a submersion.  Hence $ \mathcal{E} _h $ is a smooth manifold whose tangent space at $ \mathbf{b} $ is given by $ \ker d F _{ \mathbf{b}} $.  From \eqref{dif_FF} it is clear that, as subspaces of $ \mathbb{R} ^{ N\choose2} $, $ \ker d F _{ \mathbf{b}} = \ker d F _{ \lambda \mathbf{b} } $ for all $ \mathbf{b} \in Q $, $ \lambda > 0 $.  Therefore, 
\[
  \lim _{ t \rightarrow T _\ast} T _{ \mathbf{b} (t)} \mathcal{E} _h = \lim _{ t \rightarrow T _\ast} T _{ \hat{ \mathbf{b}} (t) } \mathcal{E} _h = T _w \mathcal{E} _h \,, 
\] 
where $ w := \lim _{ t \rightarrow T _\ast} \hat{ \mathbf{b}} (t) $ is guaranteed to exist by the assumption that $ \mathbf{b} (t) $ approaches the origin in a regular way.

On the other hand, from lemma \ref{lem:regular_curve} and the remark following definition \ref{def:orig_reg_way}, $ w = \pm \lim _{ t \rightarrow T _\ast} \hat{ \dot{b} } (t) $.  Since $ \hat{ \dot{b} } (t) \in T _{ b(t)} \mathcal{E} _h $ for all $t$, we have that $ w \in T _w \mathcal{E} _h $.  Hence,
\[
  0 = d F _w \cdot w = F(b) (\Gamma _1 \Gamma _2 + \cdots + \Gamma _{ (N-1)} \Gamma _N ) \,. 
\] 
Thus $ V = 0 $, as claimed.
\end{proof}%

\subsection{Application: total collision for a vortex parallelogram}
\leftlabel{s-VPLG}

In this section we consider a system of four point-vortices in parallelogram configuration with the property that the vorticities at opposite vertices are equal.  (This problem is discussed in \cite{NovikovSedov1979} assuming from the start a self-similar collision.)  We will use proposition \ref{prop:reg_collapse_virial} to show that the system admits total collapse only if $ V = 0 $.

\todo{A: Mention that Novikov has considered the parallelogram problem and compare with our discussion.}

Consider then a four point-vortex with $ \Gamma _1 = \Gamma _3 $, $ \Gamma _2 = \Gamma _4 $ and an initial condition satisfying $ b _{ 12} = b _{ 34}, b _{ 13} = b _{ 24} $.  Then, from \ref{sqdist_dynecs}, it follows that
\[
  \dot{b} _{ ij} = \frac{A _{ ijk} }{ 2 \pi} \left( \frac1{b _{ ik}} - \frac1{b _{ jk}} \right) \left( \Gamma _i - \Gamma _j \right) \,, 
\] 
with $ (i,j,k,l) $ any permutation of $ (1,2,3,4) $, and hence $ \dot{b} _{ 12} - \dot{b} _{ 34} = \dot{b} _{ 13} - \dot{b} _{ 24} = 0 $.  This shows that the property of being in a parallelogram configuration is preserved through out the evolution of the system.

Now let us consider evolutions leading to total collapse.  Then we have to assume that $ M = 0 $ and $ \Gamma _1 \Gamma _2 < 0 $.  The conservation of energy $H$ and inertia $M$, together with the \emph{parallelogram law}
\[
  b _{ 12} + b _{ 34} = 2(b _{ 13} + b _{ 14})
\] 
imply that 
\begin{equation} \label{sys:parallelogram_curve1}
  \left| \Gamma _1 \Gamma _2 \right| \left( b _{ 13} + b _{ 34} \right) = \Gamma _1 ^2 \, b _{ 13} + \Gamma _2 ^2 \, b _{ 24} \,, \quad
  b _{ 13} ^{ \Gamma _1 ^2 } \, b _{ 24} ^{ \Gamma _2 ^2} = h \left( b _{ 12} \, b _{ 14} \right) ^{ 2 | \Gamma _1 \Gamma _2 |} \,. 
\end{equation}
Noticing that the first equation in \eqref{sys:parallelogram_curve1} implies that $ b _{ 13} = | \Gamma _2 / \Gamma _1 | b _{ 24} $, it follows that system \eqref{sys:parallelogram_curve1} is equivalent to
\begin{equation} \label{sys:parallelogram_curve2}
	f _1 (\beta) z = h\, (x\, y) ^\delta \,, \quad (1+\beta)z = 2(x + y) \,,
\end{equation}
where $ (x,y,z) = (b _{ 12}, b _{ 14}, b _{ 24}) $, $ \beta = | \Gamma _2 / \Gamma _1 | $, $ \delta = 2 | \Gamma _1 \Gamma _2 | / (\Gamma _1 ^2 + \Gamma _2 ^2) $ and $ f _1 (x) = x ^{ 1/(1+x ^2)} $.
Eliminating $z$ from \eqref{sys:parallelogram_curve2} we obtain
\[
  f _2 (\beta ; h) (x+y) = (x\, y) ^\delta \,, 
\] 
where $ f _2 (x; h) = 2 f _1(x)/(h(1+x)) $.  Thus, introducing the \emph{projective} coordinate $ p = y/x $ one gets the following reparametrization of the evolution curve of $ (x(t), y(t), z(t)) $, 
\begin{equation} \label{parallelogram_paramcurve1}
	x = \left( \frac{ \gamma\, p ^\delta }{ 1 + p} \right) ^\alpha \,, \quad 
	y = \left( \frac{ \gamma\, p ^{(1- \delta)} }{ 1 + p} \right) ^\alpha \,, \quad 
	z = A \left( \frac{ \gamma\, p }{ 1 + p} \right) ^\alpha \,, 
\end{equation}
where $ A = h/f _1 (\beta) $, $ \gamma = f _2 (\beta ; h) ^{-1} $ and $ \alpha = 1/(1-2 \delta) $.  An analogous curve is obtained using the projective coordinate $ q = x/y $ as parameter.  Notice that $ 1-2 \delta = (| \Gamma _1 | + | \Gamma _2 |) ^2 / ( | \Gamma _1 | ^2 + | \Gamma _2 | ^2 ) > 1 $.  Hence $ 0 < \alpha < 1 $. 

We are interested in the case when $ (x,y,z) \rightarrow \mathbf{0} $ as $ p \rightarrow 0 $ (or $ q \rightarrow 0 $).  This implies that $ 0 < \delta < 1 $, which we now assume in what follows.  For $ p \neq 0 $, $ (x(p), y(p), z(p)) ^\wedge = \hat{ c} (p) $ where 
\[
  c(p) = \left( p ^{ \alpha \delta}, p ^{ \alpha (1- \delta)}, A \, p ^\alpha \right) \,. 
\] 
Dividing $ c(p) $ by $ p ^{ \alpha \delta} $ (by $ p ^{ \alpha(1- \delta)} $) if $ 0 < \delta \le 1/2 $ (if $ 1/2 \le \delta < 1 $) we obtain that $ \lim _{ p \rightarrow 0} \hat{ c}(p) $ equals $(1,0,0)$, $(1,1,0)$ or $(0,1,0)$ depending on whether $ 0 < \delta < 1/2 $, $ \delta = 1/2 $ or $ 1/2 < \delta < 1 $, respectively.  Hence $ (x(p), y(p), z(p)) $ is a curve that approaches the origin in a regular way as $ p \rightarrow 0 $.  It is clear that an analogous conclusion is reached if we work with the parameter $ q = 1/p $ instead.  Therefore, by proposition \ref{prop:reg_collapse_virial}, the four-vortex problem in parallelogram configuration under consideration admits total collapse only if $ V = 0 $.

\section{Necessary properties of collision evolutions in bounded dynamics}
\label{sec:candidate collapse configs}

In this section we establish some necessary properties that any candidate dynamic evolution, leading to binary or ternary collisions in the 4-vortex problem, must satisfy.  The properties to be discussed refer to the distances between the various elements of the limit configuration (as the time tends to the collision time) and necessary relations to be satisfied by the vorticities.  These properties are obtained only from energy and inertia considerations.

\leftlabel{s-LAC1}

\note{A: Maybe should point out that this discussion is motivated by conclusions, based on the energy equation, for bounded dynamics.}

The energy equation \eqref{energy_1} alone restricts the types of collisions that can occur if we assume that the system remains bounded for all time.  Indeed,

\begin{proposition}
	If a system of four point vortices remains in a bounded domain while it evolves towards a collision then it is either a ternary collision, a double binary collision or a total collision.
\end{proposition}

\begin{proof} 
The energy equation \ref{energy_2}
implies that if the system is bounded and develops a collision then not all the vorticities have the same sign.  W.l.o.g. we can assume that either a) $ \Gamma _1, \Gamma _2 > 0 $ and $ \Gamma _3, \Gamma _4 < 0 $ or b) $ \Gamma _1, \Gamma _2, \Gamma _3 > 0 $ and $ \Gamma _4 < 0 $.

\paragraph{Case (a).}  Equation \eqref{energy_2} becomes
\begin{equation} \label{energy_2-caseA}
	b _{ 12} ^{ | \Gamma _1 \Gamma _2 |} b _{ 34} ^{ | \Gamma _3 \Gamma _4 |} = h\, b _{ 13} ^{ | \Gamma _1 \Gamma _3 |} b _{ 14} ^{ | \Gamma _1 \Gamma _4 |} b _{ 23} ^{ | \Gamma _2 \Gamma _3 |} b _{ 24} ^{ | \Gamma _2 \Gamma _4 |} \,. 
\end{equation}
If one of the factors in the l.h.s. of \eqref{energy_2-caseA} tends to zero (two vortices of the same sign collide) then at least one of the terms in the r.h.s. tends to zero (two vortices of opposite sign collide).  This leads to a ternary collision.  Similarly, collision of two vortices of opposite sign implies collision of two vortices of the same sign and again we have a ternary collision.  Notice that, if $ M = 0 $, the ternary collision can be part of a total collision.

\paragraph{Case (b).}  Equation \eqref{energy_2} becomes
\begin{equation} \label{energy_2-caseB}
	b _{ 12} ^{ | \Gamma _1 \Gamma _2 |} b _{ 13} ^{ | \Gamma _1 \Gamma _3 |} b _{ 23} ^{ | \Gamma _2 \Gamma _3 |} = h\, b _{ 14} ^{ | \Gamma _1 \Gamma _4 |} b _{ 24} ^{ | \Gamma _2 \Gamma _4 |} b _{ 34} ^{ | \Gamma _3 \Gamma _4 |} \,. 
\end{equation} 
Suppose that two vortices of the same sign collide, say $ b _{ 12} \rightarrow 0 $.  Then at least one of the factors in the r.h.s. of \eqref{energy_2-caseB} must tend to zero.  There are two distinct possibilities, represented by $ b _{ 14} \rightarrow 0 $ and by $ b _{ 34} \rightarrow 0 $.  (The case $ b _{ 24} \rightarrow 0 $ represents the same case as $ b _{ 14} \rightarrow 0 $.)  The first possibility leads to the ternary collision $ (1,2,4) $ and the second leads to the double binary collision $ (1,2), (3,4) $.  If we start by assuming that two vortices of opposite sign collide then at least one of the factors in the l.h.s. of \eqref{energy_2-caseB} must tend to zero and the same argument applies.
\end{proof} 

\begin{remark}
	In the context of the proof of the previous proposition, 
	it is easy to check that in case (a) (i.e. $ \Gamma l1, \Gamma _2 > 0 $ and $ \Gamma _3, \Gamma _4 < 0 $) there are four ways in which the vortices can conglomerate in a candidate proper ternary collision, namely, one for each vortex excluded from the collision of the other three.  And in case (b) (i.e. $ \Gamma _1, \Gamma _2, \Gamma _3 > 0 $ and $ \Gamma _4 < 0 $) there are three possibilities for a proper ternary collision, namely one for each positive vortex excluded, and there are three possibilities for a double binary collision, namely one for each of the positive vortices colliding with the negative one while the other two positive vortices collide.
\end{remark}

\subsection{Conditions and configurations for candidate partial collisions}

Let us consider what are the implications of conservation of $M$ for any candidate evolution leading to a (not necessarily proper) triple collision or double binary collision.  All vortices in the system are assumed to have non-zero vorticity.

\paragraph{Ternary collision.}  Suppose that there is an evolution leading to the collapse of vortices $ (1, 2, 3) $.  Then $ b _{ 14} - b _{ 24} \rightarrow 0 $ and $ b _{ 14} - b _{ 34} \rightarrow 0 $. Therefore \eqref{Meq} implies that, in the limit as time approaches the collision time, 
\begin{equation} \label{M_balance1}
	M = ( \Gamma _1 + \Gamma _2 + \Gamma _3 ) \Gamma _4 d ^2 
\end{equation}
where $ d ^2 = b _{ 14} = \| z _{ 123} ^\ast - z _4 ^\ast \| ^2 $.  Here $ z _{ 123} ^\ast , z _r ^\ast $ are the positions of the colliding point of vortices $ (1,2,3) $ and of the limit point of vortex $4$.

\paragraph{Double binary collision.}  Suppose that there is an evolution leading to the simultaneous collapse of vortices $ (1,2) $ and $ (3,4) $.  Then \eqref{Meq} implies that, in the limit as time approaches the collision-time, 
\begin{equation}\label{M_balance2}
	M = (\Gamma _1 + \Gamma _2) (\Gamma _3 + \Gamma _4) d ^2 \,, 
\end{equation} 
where $ d ^2 = b _{ 13} = \| z _{ 12} ^\ast, z _{ 34} ^\ast \| ^2 $.  Here $ z _{ 12} ^\ast, z _{ 34} ^\ast $ are the positions of the colliding points of vortices $ (1,2) $ and $ (3,4) $.

From the discussion of the cases leading to equations \eqref{M_balance1} and \eqref{M_balance2} we arrive at the following collision conditions depending on whether M is equal to zero or not, on the vorticities and on the limit configuration at collision time:

\begin{proposition} 
Consider a system of four point-vortices each with non-zero vorticity.  Use the notation $ \Gamma _{ ij} = \Gamma _i + \Gamma _j $ and $ \Gamma _{ ijk} = \Gamma _i + \Gamma _j + \Gamma _k $.  Then necessary conditions for ternary or double binary collapse are given by the following table:
\vspace{1ex}
\begin{center} \begin{tabular}{|c|c|c|c|}
	\hline
	\parbox{17ex}{\begin{center}
		Collapse type / vortices colliding
	\end{center}} & Condition on $M$ & Vorticities & \parbox{16.5ex}{\begin{center}
		Limit configura-\\tion parameter
	\end{center}} \\
	\hline\hline
	\multirow{2}{*}{\parbox{9ex}{\begin{center}
		Ternary\\$ (i,j,k) $
	\end{center}}} & if $ M = 0 $ & $ \Gamma _{ ijk} = 0 $ & $d$ arbitrary \\
	\cline{2-4}
	& if $ M \neq 0 $ & \parbox{14ex}{\begin{center}
		$ \Gamma _{ ijk} \neq 0 $, \\ $ M \, \Gamma _{ ijk} \, \Gamma _l > 0 $
	\end{center}} & $ d = \sqrt{\frac{ M }{ \Gamma _{ ijk} \Gamma _l }} $ \\
	\hline
	\multirow{2}{*}{\parbox{15ex}{\begin{center}
		Double binary \\ $ (i,j) $ and $ (k,l) $ 
	\end{center}}} & if $ M = 0 $ & $ \Gamma _{ ij} \Gamma _{ kl} = 0  $ & $d$ arbitrary \\
	\cline{2-4}
	& if $ M \neq 0 $ & \parbox{14ex}{\begin{center}
		$ \Gamma _{ ij} \Gamma _{ kl} \neq 0 $, \\ $ M \, \Gamma _{ ij} \,\Gamma _{ kl} > 0 $ 
	\end{center}} & $ d = \sqrt{\frac{ M }{ \Gamma _{ ij} \Gamma _{ kl} }} $ \\
	\hline
\end{tabular} \end{center}
\vspace{1ex}
Here $d$ is the distance between the limit cluster points, as time approaches the collision time.
\end{proposition} 

\section{On the relationship between partial and total collapse}
\label{sec:rel_par_tot_collapse}

\leftlabel{s-BTC}

In this section we drop the assumption of bounded dynamics.  In the proposition and corollary that follows we consider the setting of a system o four point vortices, all with non-zero vorticities and such that two of them, say $ \Gamma _1 $ and $ \Gamma _2 $, are of opposite sign.

\note{A: Maybe should motivate the assumptions of the following proposition.  The assumption on the vorticities is generically true;  the assumption on $M$ is true for a system that is candidate to total collision.}

\leftlabel{t-BTC1}

\begin{proposition}
	Suppose that $ (\Gamma _1 + \Gamma _2) (\Gamma _1 + \Gamma _2 + \Gamma _3) \cdot (\Gamma _1 + \Gamma _2 + \Gamma _4 ) \neq 0 $ and $ M = 0 $.  Let $ \{ \beta _{ ij} \} $ be the normalized 
\todo{A:  Define 'normalized' square distances and give reference.  Also define the 'quadrilateral scale factor'.} \noindent%
square distances between the vortices, as defined above.  Suppose that, as $ t \rightarrow T _0 \le \infty $, $ \beta _{ 12} \rightarrow 0 $.  Then there is a $ T < T _0 $ and there are constants $ 0 < c _1 < c _2 $ such that if $ T < t < T _0 $ then $ c _1 < \beta _{ 12} ^{ | \Gamma _1 \Gamma _2 | } / \rho ^V < c _2 $, where $V$ is the virial and $\rho$ is the quadrilateral scale factor defined above.
\end{proposition}

\proof Noticing that $ \beta _{ 12} \rightarrow 0 $ implies that $ \beta _{ 23} - \beta _{ 13} \rightarrow 0 $ and $ \beta _{ 24} - \beta _{ 14} \rightarrow 0 $ we obtain, from \eqref{energy_2}, that
\[
  \frac{ h \, \beta _{ 12} ^{ | \Gamma _1 \Gamma _2 |} }{ \rho ^V } = \beta _{ 13} ^{ \Gamma _{ 12} \Gamma _3} \beta _{ 14} ^{ \Gamma _{ 12} \Gamma _4 } \beta _{ 34} + \varphi _1 (t) \,, 
\] 
where $ \Gamma _{ 12} := \Gamma _1 + \Gamma _2 $.  The constraint $ \sum \beta _{ ij} = 1 $ implies that
\begin{equation} \label{normalization_beta}
  2 \beta _{ 13} + 2 \beta _{ 14} + \beta _{ 23} = 1 + \varphi _2 (t) \,. 
\end{equation}
The condition $ M = 0 $ 
\todo{A: State and give reference to condition `M=0'.} \noindent%
implies
\begin{equation} \label{M 3vor}
  \Gamma _{ 12} \Gamma _3 \beta _{ 13} + \Gamma _{ 12} \Gamma _4 \beta _{ 14} + \Gamma _3 \Gamma _4 \beta _{ 34} = \varphi _3 (t) \,. 
\end{equation}
Here the functions $ \varphi _i (t) $, $ i = 1, 2, 3 $, tend to zero as $ t \rightarrow T _0 $.  Since the triple $ ( \beta _{ 13}, \beta _{ 14}, \beta _{ 34} ) $ represent the squares of the sides of a triangle then it lies in the cone 
\begin{equation} \label{cone ineq}
  \beta _{ 13} ^2 + \beta _{ 14} ^2 + \beta _{ 34} ^2 \le 2 \left( \beta _{ 13} \beta _{ 14} + \beta _{ 13} \beta _{ 34} + \beta _{ 14} \beta _{ 34} \right) \,. 
\end{equation}
The intersection $\mathcal{L}$ of the line given by the limits, a $ t \rightarrow T _0 $, of \eqref{normalization_beta} and \eqref{M 3vor} with the cone \eqref{cone ineq}, if not empty, is a closed 
segment and thus a compact set.  (If $ \mathcal{L} = \emptyset $ then $ \beta _{ 12} $ can not tend to zero and the proposition is trivially true.)  Identify $ ( \beta _{ 13}, \beta _{ 14}, \beta _{ 34} ) $ with  point $  (x, y, z) \in \mathbb{R} ^3 $.  The cone \eqref{cone ineq} is tangent to the cartesian planes at the lines $ (x = y, z = 0 ) $, $ (x = z, y = 0 $, $ (y = z, x = 0 ) $.  The hypothesis  $ (\Gamma _1 + \Gamma _2) (\Gamma _1 + \Gamma _2 + \Gamma _3) \cdot (\Gamma _1 + \Gamma _2 + \Gamma _4 ) \neq 0 $  implies that these lines do not intersect \eqref{M 3vor} in the limit as $ t \rightarrow T _0 $.  Therefore $ x\, y\, z \neq 0 $ for all $ (x,y,z) \in \mathcal{L} $.  Let
\[
  ( \bar{ c} _1, \bar{ c} _2) = \mathop{(\min, \max)}_{(x,y,z) \in \mathcal{L}} x ^{ \Gamma _{ 12} \Gamma _3 }\, y ^{ \Gamma _{ 12} \Gamma _4}\, z ^{ \Gamma _3 \Gamma _4} \,. 
\] 
Since the distance between the point $ (x,y,z) $ and the segment $\mathcal{L}$ tends to zero as $ t \rightarrow T_0 $ then, for any $ \delta >0 $, there exists $ T < T _0 $ such that if $ T < t < T _0 $ then $ \bar{ c} _1 - \delta < x ^{ \Gamma _{ 12} \Gamma _3 }\, y ^{ \Gamma _{ 12} \Gamma _4}\, z ^{ \Gamma _3 \Gamma _4} < \bar{ c} _2 + \delta $.  Choosing $ (c _1, c _2) = (\bar{ c} _1 - \delta, \bar{ c} _2 + \delta)/h $, with $ \delta = \bar{ c} _1 / 2 $, we get the desired result. \qed

\leftlabel{t-BTC2}

\begin{corollary} \label{cor:partial->total collapse}
	Suppose that vortices $1$ and $2$ have vorticities of opposite sign, that $ (\Gamma _1 + \Gamma _2)(\Gamma _1 + \Gamma _2 + \Gamma _3)(\Gamma _1 + \Gamma _2 + \Gamma _4) \neq 0 $ and that $ M = 0 $.  Then, if $ V > 0 $ (resp. $ V < 0 $), $ \beta _{ 12} \rightarrow 0 $ only if $ \rho \rightarrow 0 $, (resp. $ \rho \rightarrow \infty $).  Thus, a relative binary or ternary collision involving vortices $1$ and $2$ 
can occur only as part of a process of total collision if $ V > 0 $ and as part of a blow-up process if $ V < 0 $.
\end{corollary}

\note{A: Maybe we can remark that the previous corollary can be interpreted as saying that partial collision occurs only as part of a process of total collapse or blow-up.}

\leftlabel{t-TCiBC1}

\begin{proposition}
	Suppose that a system of four point vortices, all with non-zero vorticity, evolves (maybe sequentially) towards total collision.  Then the system evolves towards a binary or ternary sequential collision.
\end{proposition}

\begin{proof} 
If the system evolves towards total collision then there is a sequence of times $ \{ t _n \} $, $ t _n \rightarrow T _\ast \le \infty $, such that $ \{ \rho (t _n ) \} $ is a sequence converging to zero.  It follows from \eqref{energy_1} that if $ V > 0 $ (resp. $ V < 0 $) then there must be at least one pair of labels $ (i,j) $ with $ \Gamma _i \Gamma _j < 0 $ (resp. $ \Gamma _i \Gamma _j > 0 $) such that $ \{ \beta _{ ij} (t _{ n _m}) \} $ is a sequence converging to zero, with $ \{ t _{ n _m} \} $ a subsequence of $ \{ t _n \} $.  This implies that vortices $i$ and $j$ evolve weakly towards collision.
\end {proof}  

\section{Dynamics of a four-vortex system with a binary}

We have seen that, when the virial is positive, total collapse has to be accompanied by a binary or ternary collapse.
\todo{A: Actually I have only showed this for the three positive and one negative vorticity; I am assuming that the same happens with two positive and two negative vorticities. Remains to be verified.}\noindent 
In this section we consider the dynamics of a four-point vortex system in a state very close to having a binary collision.  The two vortices that are near collision can be called a \textbf{vortex binary}.

Consider then a system of four point vortices with $ \beta _{ 14} << 1 $.  Let its configuration be given by $ (z _1,z  _2,z _3, z _4) \in \mathbb{C} ^4 $, so that $ |z _1 - z _4 | $ is very small relative to (some of) the other distances.  Introduce the linear change of coordinates 
\[
  z _1 = \zeta - \frac{ \Gamma _4 }{ \Gamma _{ 14}} q \,, \quad z _2 = z _2 \,, \quad z _3 = z _3 \,, \quad z _4 = \zeta + \frac{ \Gamma _1 }{ \Gamma _{ 14}} q \,, 
\] 
with $ \Gamma _{ 14} := \Gamma _1 + \Gamma _4 $, $ q, \zeta \in \mathbb{C} $.  Thus $q$ is the vector going from $ z _1 $ to $ z _4 $ and $\zeta$ is the position vector of the center of vorticity of the pair $ (1,4) $.  A calculation shows that the vector field in the $ (q, \zeta, z  _2, z _3) $ coordinates is given by
\[\begin{split}
  \dot{ q} &= \frac{ \mi }{ 2 \pi } \left[ \Gamma _{ 14} \frac{ q }{ |q| ^2 } + \Gamma _2 \left( R _2 ^1 - R _2 ^4 \right) + \Gamma _3 \left( R _3 ^1 - R _3 ^4 \right) \right] \,, \\
  \dot{ \zeta} &= \frac{ \mi }{ 2 \pi \, \Gamma _{ 14}} \left[ \Gamma _2 \left( \Gamma _1 R _2 ^4 + \Gamma _4 R _2 ^1 \right) + \Gamma _3 \left( \Gamma _1 R _3 ^4 + \Gamma _4 R _3 ^1 \right) \right] \,, \\
  \dot{ z} _2 &= \frac{ \mi }{ 2 \pi} \left[ \Gamma _3 \frac{ z _2 - z _3 }{ | z _2 - z _3 | ^2 } - \Gamma _4 R _2 ^1 - \Gamma _1 R _2 ^4 \right] \,, \\
  \dot{ z} _3 &= \frac{ \mi }{ 2 \pi} \left[ \Gamma _2 \frac{ z _3 - z _2 }{ | z _3 - z _2 | ^2 } - \Gamma _4 R _3 ^1 - \Gamma _1 R _3 ^4 \right] \,, 
\end{split}\] 
where $ \Gamma _{ 14} := \Gamma _1 + \Gamma _4 $ and
\[
  R _j ^1 = \frac{ \zeta - z _j + \frac{ \Gamma _1 }{ \Gamma _{ 14}} q }{ | \zeta - z _j + \frac{ \Gamma _1 }{ \Gamma _{ 14}} q | ^2 } \,, \quad R _j ^4 = \frac{ \zeta  z _j - \frac{ \Gamma _4 }{ cg _{ 14}} q }{ | \zeta - z _j - \frac{ cg _4 }{ \Gamma _{ 14}} q | ^2 } \,, 
\] 
$ j = 2, 3 $.

These can be regarded as the dynamic equations of two coupled systems, $ \{q \} $ and $ \{\zeta, z  _2, z _3 \} $.  Notice that, while at $ q = 0 $ the dynamic equation for $ \{q\} $ becomes singular, the dynamic equations for $ \{\zeta , z  _2,z _3 \} $ become the equations of a standard three-vortex system with vorticities $ ( \Gamma _1 + \Gamma  _4,\Gamma  _2,\Gamma _3 ) $.

Let us now consider closely the system $ \{q\} $.  With $ w _j = \zeta - z _j $, $ j = 2, 3  $, the terms $ (R _j ^1 - R _j ^4 ) $ appearing in the expression for $ \dot{q} $ can be written as 
\begin{equation}\label{R1-R4}
	R _j ^1 - R _j ^4 = \frac{ |w _j | ^2 + \frac{ \Gamma _1 \Gamma _4 }{ \Gamma _{ 14}} |q| ^2 }{ | w _j + \frac{ \Gamma _1 }{ \Gamma _{ 14}} q | ^2  | w _j - \frac{ \Gamma _4 }{ \Gamma _{ 14}} q | ^2 } \, q + \frac{ \frac{ \Gamma _4 - \Gamma _1 }{ \Gamma _{ 14}} |q| ^2 - 2 w _j \cdot q }{ | w _j + \frac{ \Gamma _1 }{ \Gamma _{ 14}} q | ^2  | w _j - \frac{ \Gamma _4 }{ \Gamma _{ 14}} q | ^2 } \, w _j \,, 
\end{equation} 
where $ w _j \cdot q := Re(w _j \bar{q}) $.  Thus,
\begin{equation} \label{qdot complete}
	\dot{q} = \frac{ \mi }{ 2 \pi} \left[ \left( \frac{ 1 }{ |q| ^2} + f _1 (w  _2,w  _3,q) \right)q + f _2 (w  _2,w  _3,q) w _2 + f _3 (w  _2,w  _3,q) w _3 \right]
\end{equation} 
with $  f_1 (w  _2,w  _3,0) = \Gamma _2/ |w _2 | ^2 + \Gamma _3 / |w _3 | ^2 $, $ f _2(w  _2,w _3, 0) = f _3 (w  _2,w  _3,0) = 0 $.
We can regard \eqref{qdot complete} as a perturbation of
\begin{equation}\label{qdot-unperturbed}
  \dot{q} = \frac{ \mi }{ 2 \pi} \left( \frac{ 1 }{ |q| ^2 } + f _1 (w  _2, w _3, q) \right) q 
\end{equation}
whose phase portrait consists of concentric circles around the origin.

We now study the change in the phase portrait due to the perturbation term $ \mi (f _2 w _2 + f _3 w _3 )/ (2 \pi)  $.  To this effect we need to understand some general aspects of the relationship between the dynamics of $ \{q\} $ and $ \{ \zeta,z  _2, z _3 \} $.

The first observation is that the total energy $h$ is approximately the product\footnote{%
We talk about the \emph{product} and not the \emph{sum} because $h$ is obtained from the exponentiation of the original energy introduced in \eqref{energy_1}.}
of the energies of $ \{q\} $ and $ \{ \zeta,z  _2,z _3 \} $.  More precisely, given $ (z _1, \ldots, z _3) \in \mathbb{C} ^n $, let $ h(z  _1, \ldots, z _n; \Gamma _1, \ldots, \Gamma _n) $, $ M(z  _1, \ldots, z _n; \Gamma _1, \ldots, \Gamma _n) $ denote the associated energy and the value of $ \sum _{ i < j} \Gamma _i \Gamma _j | z _i - z _j | ^2 $, for the given vorticities $ (\Gamma _1, \ldots, \Gamma  _n)$.  It is easy to see that if $ |q| << q $ (i.e. the binary is very small) then
\[
  h(z  _1, z  _2, z _3, z _4; \Gamma _1, \Gamma _2, \Gamma _3, \Gamma _4) \approx h(\zeta, z  _2,z _3; \Gamma _{ 14}, \Gamma _2, \Gamma _3) |q| ^{ 2 \Gamma _1 \Gamma _4} \,. 
\] 

The second observation is that, if the vorticities satisfy certain inequality, then for any given energy, if the binary is small enough, the dynamics of $ \{ \zeta, z_2, z _3 \} $ is much slower than the dynamics of $ \{q\} $.  More precisely,
\begin{proposition}\label{prop:rapid-slow dyn}
	If $ V > 2 | \Gamma _1 \Gamma _4 | $ then given $ h _0 > 0 $ and $ \varepsilon > 0 $ there exists $ \delta > 0 $ such that if $ |q| < \delta $ then, for any $ ( \zeta,z _2, z _3)  \in \mathbb{C} ^2  $ such that $ h(\zeta, z _2, z _3; \Gamma _{ 1}, \Gamma _2, \Gamma  _3)= h _0 /|q| ^{ 2 \Gamma _1 \Gamma _4}  $ and $ M( \zeta, z _2,z _3; \Gamma _{ 14}, \Gamma _2, \Gamma _3) = 0 $, we have that 
	\[
  		\frac{ | \dot{ w} _2 | + | \dot{ w} _3 | }{ | \dot{ q} | } < \varepsilon \,. 
	\] 
\end{proposition}

\proof[Sketch of proof]  Let $ \Gamma (t) = ( b _1(t), b _2(t), b _3(t)) $, $ b _1 = |z _2 - z _3 | ^2 $, $ b _2 = | \zeta - z _3 | ^2 $, $ b _3 = | \zeta - z _2 | ^2 $.  We know that, denoting $ \dot{\gamma} (t) = \dot{\gamma} _R (t) + \dot{\gamma} _B (t) $, where $ \dot{\gamma} _R $ is the radial component of the velocity and $ \dot{\gamma} _B $ is the component parallel to the plane $ B: b _1 + b _2 + b _3 = 1 $, $ \dot{\gamma} _B (t) = v _0 / \| \Gamma(t) \| $ for some constant vector $ v _0 $ parallel to $B$.  Also, if $ \rho _{\text{min}}(h) $ denotes the minimum value of $ \| \gamma(t) \|  $, with $ \gamma (t) $ constrained to have energy $h$ and $ M = 0  $, then $ \rho _{\text{min}} (h(\zeta, z  _2, z _3; \Gamma _{ 14}, \Gamma  _2, \Gamma _3)) \sim h(\zeta, z  _2, z_3; \Gamma _{ 14}, \Gamma _2, \Gamma _3) ^{ 1/V} = h _0 |q| ^{ -2| \Gamma _1 \Gamma _4 |/V} $.  On the other hand, $ | \dot{q} | \sim 1/ |q| $.  Therefore $ | \dot{\gamma} _B | / | \dot{q} | \sim |q| ^{ 1 - 2 | \Gamma _1 \Gamma _4 | / V } $ and the result follows. \qed 

\todo{A: This ``sketch of proof'' still needs to be reviewed carefully and needs to be replaced by an argument considering $ \dot{ w} _j  $ ($ j = 2, 3 $) directly.}

\begin{remark} 
We have seen above that it can happen that, besides $ \beta _{ 14} $,  $ \beta _{ j4} $ tends to zero as $ t \rightarrow t _0 $ for $ j = 2 $ or $3$ (but not both).  If such is the case then we have a ternary collision and thus $ |q| $ and $ | w _j | $ may be comparable.  One can study the case when this is not the case, so that $ \Gamma _4 $ participates only in a binary collision.  More precisely, one can assume that $ |q| ^2 / | w _j | ^3 << 1 $, $ j = 2, 3 $.  In this case the denominator in both terms of the r.h.s. of \eqref{R1-R4} can be approximated by $ | w _j | ^4 $ and the perturbation term can be approximated as
\[
  \frac{ \mi }{ 2 \pi} (f _2 w _2 + f _3 w _3) \approx \frac{ \mi }{ 2 \pi} \left[ \frac{ \Gamma _{ 124} |q|^2 - 2 w _2 \cdot q }{ | w _w | ^4} \, w _2 + \frac{ \Gamma _{ 134} |q| ^2 - 2 w _3 \cdot q }{ | w _3 | ^4} \, w _3 \right] \;, 
\] 
with $ \Gamma _{ 1j4} := \Gamma _j (\Gamma _4 - \Gamma _1)/(\Gamma _1 + \Gamma _4) $ and with
\begin{equation} \label{perturb wj small}
  \frac{ \Gamma _{ 1j4} |q| ^2 - 2 w _j \cdot q }{ | w _j | ^4} \, w _j << 1 \;, \quad j = 2, 3 \;. 
\end{equation}
\end{remark} 

\subsection{A canonical transformation allowing averaging}

We have seen above that the equations of motion for the four vortex problem with a binary can be thought of two coupled systems, one for the binary and one for a three-vortex system, with an added perturbation that couples the system.  In this subsection we will discuss a canonical transformation that allows to formally apply the averaging principle (cf. \cite[chap. 10, \S 52]{Arnold1978} or \cite{SandersVerhulst1985}) in order to simplify the effect of the perturbation on the three-vortex system.  In future work we will deal with the delicate problem of determining the precise conditions under which the averaged dynamics is a good approximation for asymptotically  large times as the binary becomes small.

Let $ \{ z _i ; \Gamma _i \} $, $ i = 1, 2, 3, 4 $ be the complex coordinates and vorticities of a system of four point vortices such that initially the distance between vortices $ z _1 $ and $ z _2 $ is very small.  The symplectic matrix in coordinates $ (x _i, y _i ) $, with $ z _i = x_i + \mi y _i $, that corresponds to \eqref{symplectic_form}, is given by
\[
  \mathbb{J} _0 = \begin{pmatrix} 0 & J _0 \\ - J _0 & 0 \end{pmatrix}\,, \text{ with } J _0 = \begin{pmatrix} \Gamma _1 & & & \\ & \Gamma _2 & & \\ & & \Gamma _3 & \\ & & & \Gamma _4 \end{pmatrix} \,. 
\] 
We start with a transformation $ T _1 $ that takes the original complex coordinates to the coordinates $ \{\zeta, z, z _3, z _4 \} $, where $\zeta$ is the vector from $ z _1 $ to $ z _2 $ and $z$ denotes the position of the center of vorticity of the binary system $ (z _1, z _2 ) $.  Then $ T _1 $ is the linear transformation given by the matrix $ \mathcal{R} (\tilde{ T} _1) $, where $ \mathbb{R} : \operatorname{ GL}(n, \mathbb{C}) \longrightarrow \operatorname{ GL} (2n, \mathbb{R}) $ is the operator transforming a complex matrix into its real form and  
\[
  \tilde{ T} _1 = \begin{pmatrix} -1 & 1 & 0 & 0 \\ \frac{ \Gamma _1 }{ \Gamma _1 + \Gamma _2 } & \frac{ \Gamma _2 }{ \Gamma _1 + \Gamma _2 } & 0 & 0 \\ 0 & 0 & 1 & 0 \\ 0 & 0 & 0 & 1 \end{pmatrix} \,. 
\] 
A calculation shows that $ T _1 ^t\, \mathbb{J} _0\, T _1 = \mathbb{J} _1 $ with
\[
  \mathbb{J} _1 = \begin{pmatrix} 0 & J _1 \\ -J _1 & 0 \end{pmatrix} \text{ and } J _1 = \begin{pmatrix} \frac{ \Gamma _1 \Gamma _2 }{ \Gamma _1 + \Gamma _2} & & & \\ & \Gamma _1 + \Gamma _2 & & \\ & & \Gamma _3 & \\ & & & \Gamma _4 \end{pmatrix} \,, 
\] 
so that $ T _1 $ is a generalized canonical transformation.

The coordinates in the target space of $ T _2 $ already correspond to the two systems in which we want to split the four-vortex problem.  However, the dynamics in in the 3-vortex system needs to be reduced in order to simplify the description of the dynamics.  To this end, following \cite{ArefPomphrey1980}, we introduce a discrete Fourier transform given by:
\[
  q _n + \mi p _n = N ^{ -1/2} \sum _{ \alpha = 1} ^N z _\alpha Exp[\mi 2 \pi n ( \alpha -1) / N ]
\] 
applied to $ \{z, z _3, z _4 \} $ with $ N = 3 $.  The resulting linear transformation, taking the canonical coordinates $ \{x, x _0, x _3, x _4, y, y _0, y _3, y _4 \} $ to the new canonical coordinates $ \{x, q _0, q _1, q _2, y, p _0, p _1, p _2 \} $ is given by $ T _2 = \mathcal{R} (\tilde{ T} _2 ) $ with 
\[
  \tilde{ T} _2 = \begin{pmatrix} 1 & 0 & 0 & 0 \\ 0 & 1/\sqrt3 & 1/\sqrt3 & 1/\sqrt3 \\ 0 & 1/\sqrt3 & \me ^{ 2 \mi \pi } / \sqrt3 & \me ^{ -2 \mi \pi } / \sqrt3 \\ 0 & 1/\sqrt3 & \me ^{ -2 \mi \pi } / \sqrt3 & \me ^{ 2 \mi \pi } / \sqrt3 \end{pmatrix} \,. 
\]
Let $ \mathbb{J} _2 $ be defined by $ T _2^t \, \mathbb{J} _1 \; T _2 = \mathbb{J} _2 $.  A calculation shows that, $ \mathbb{J} _2 = \begin{pmatrix} A & B \\ -B & A \end{pmatrix} $ with $A$ antisymmetric and $B$ symmetric.  In order to have a canonical transformation in the generalized sense we need to impose the condition $ A = 0 $ and that $ B $ be a diagonal matrix.  A calculation shows that this happens if and only if 
\begin{equation} \label{Pomphrey_vort_cond}
	\Gamma _1 + \Gamma _2 = \Gamma _3 = \Gamma 4 := \Gamma \,. 
\end{equation}
In what follows we assume this condition.  A calculation shows that in this case $ \mathbb{J} _2 = \mathbb{J} _1 $.  

A further reduction is achieved by introducing semi-polar coordinates.  This is a standard procedure in celestial mechanics.  The corresponding transformation taking the coordinates $ \{x, q _0, q _1, q _2, y, p _0, p _1, p _2 \} $ to the new coordinates $ \{j, j _0, j _1, j _2, \theta, \theta _0, \theta _1, \theta _2 \} $ is defined by
\begin{multline*}
  F _3(x, q _0, q _1, q _2, y, p _0, p _1, p _2) = \\
\Big( \frac{ x ^2 + y ^2 }{ 2}, \frac{ q _0 ^2 + p _0 ^2 }{ 2}, \frac{ q _1 ^2 + p _1 ^2 }{ 2}, \frac{ q _2 ^2 + p _2 ^2 }{ 2}, \\
\operatorname{ ArcTan}(y/x), \operatorname{ ArcTan}(p _0 /q _0 ), \operatorname{ ArcTan}(p _1 /q _1 ), \operatorname{ ArcTan}(p _2 /q _2 ) \Big) \,. 
\end{multline*}
A calculation shows that this is a canonical transformation with induced symplectic form $ \mathbb{J} _3 = \mathbb{J} _2 $.

A final transformation is motivated by the fact that the first term $ H _0 $ in the Taylor expansion of the hamiltonian in the variables $ \{j, j _i, \theta, \theta _i \} $, taking $j$ as the expansion parameter, depends on $ \theta _1 $ and $ \theta _2 $ only through its difference.  (Notice that $ H _0 $ can be interpreted as the hamiltonian of the 3-vortex system unperturbed by the presence of the binary-system.)  Thus we introduce the transformation $ T _4 $ induced by the generating function 
\[
  G _3 = - \varphi j - \varphi _0 j _0 + \varphi _1 (j _1 - j _2) + \varphi _2 (j _1 + j _2) \,, 
\] 
and $ \theta _i  = - \partial  G _3 / j _i \,, \quad i _i = - \partial G _3 / \partial \varphi _i $.  The induced fourth symplectic matrix defined by $ T _4 ^t \, \mathbb{J} _3 \, T _4 = \mathbb{J}_4 $ is given by 
\[
  \mathbb{J} _4 = \begin{pmatrix} 0 & J _4 \\ -J _4 \end{pmatrix} \,, \text{ with } J _4 = \begin{pmatrix} \frac{ \Gamma _1 \Gamma _2 }{ \Gamma } &&& \\ & \Gamma && \\ && \Gamma & \\ &&& \Gamma \end{pmatrix} \,,
\] 
and with $\Gamma$ as defined in \eqref{Pomphrey_vort_cond}.  The transformation $ T _4 $ takes the canonical coordinates $ \{j, j _0, j _1, j _2, \theta, \theta _0, \theta _1, \theta _2 \} $ to the new canonical coordinates $ \{i, i _0, i _1, i _2, \varphi, \varphi _0, \varphi _1, \varphi _2 \} $.

\subsection{Averaging and hamiltonian reduction}

We are now in a position to try to simplify the hamiltonian of the four-vortex problem using the canonical variables $ \{i, i _0, i _1, i _2, \varphi, \varphi _0, \varphi _1, \varphi _2 \} $.  The variable $i$ represents the square of the radius of the binary-system.  Let $ \epsilon = \sqrt{2\, i} $ and write
\[
  H = H _0 + \epsilon H _1 + \epsilon ^2 H _2 + O(\epsilon ^3) \,. 
\] 
A calculation shows that 
\begin{multline*}
  4\, \pi H _0 = 2\,\Mvariable{\Gamma_2}\,
     \big( -\Gamma  + \Mvariable{\Gamma_2} \big) \,\log (\epsilon ) - 
    {\Gamma }^2\,\big( \log (-2\,
         \big( \Mvariable{i_2} + 
           {\sqrt{-\Mvariable{i_1} - \Mvariable{i_2}}}\,
            {\sqrt{\Mvariable{i_1} - \Mvariable{i_2}}}\,
            \cos (2\,\Mvariable{\varphi_1}) \big) ) \\
            + 
       \log ({\sqrt{-\Mvariable{i_1} - \Mvariable{i_2}}}\,
          {\sqrt{\Mvariable{i_1} - \Mvariable{i_2}}}\,
          \cos (2\,\Mvariable{\varphi_1}) - 
         2\,\big( \Mvariable{i_2} + 
            {\sqrt{3}}\,{\sqrt{-\Mvariable{i_1} - \Mvariable{i_2}}}\,
             {\sqrt{\Mvariable{i_1} - \Mvariable{i_2}}}\,
             \cos (\Mvariable{\varphi_1})\,\sin (\Mvariable{\varphi_1})
            \big) ) \\
            + \log (-2\,\Mvariable{i_2} + 
         {\sqrt{-\Mvariable{i_1} - \Mvariable{i_2}}}\,
          {\sqrt{\Mvariable{i_1} - \Mvariable{i_2}}}\,
          \cos (2\,\Mvariable{\varphi_1}) + 
         {\sqrt{3}}\,{\sqrt{-\Mvariable{i_1} - \Mvariable{i_2}}}\,
          {\sqrt{\Mvariable{i_1} - \Mvariable{i_2}}}\,
          \sin (2\,\Mvariable{\varphi_1})) \big) \,. 
\end{multline*}
Notice that the first term in the r.h.s. is proportional to $ \log(\epsilon) $ and thus blows-up as $ \epsilon \rightarrow 0 $.  However, since $ H _0 $ is cyclic in $ \varphi, \varphi _2 $, we have that, in the unperturbed dynamics, $\epsilon$ is, not surprisingly, a constant parameter.

It turns out that the second term in the $\epsilon$-Taylor expansion of the hamiltonian vanishes; that is to say, $ H _1 = 0 $.  The third term, however, is an important one and has a very complicated expression.  To make it tractable we apply the averaging principle assuming that the $ \varphi $, the variable describing rotation in the binary-system, is a fast variable.  Although we discussed above conditions under which this is a reasonable assumption, a rigorous study of the time scales in which $\varphi$ can be considered a fast variable is left for a forthcoming paper.

The averaging principle is applied formally by space-averaging the hamiltonian $ H _2 $ with respect to the variable $\varphi$, i.e.
\[
  \bar{ H} _2 = \frac1{2 \pi} \int _0 ^{ 2 \pi} H _2 \, d \varphi \,. 
\] 
A calculation implemented in \mathematica\ yields that
\[
  \bar{H} _2 = \frac{ f _1 (i _1, i _2, \varphi _1) }{ f _2 (i _1, i _2, \varphi _1 ) }
\] 
where 
\begin{multline*}
  f _1 := \big( -1 + 2\,\pi  \big) \,{\Mvariable{\Gamma_2}}^2\,
  \big( -\Gamma  + \Mvariable{\Gamma_2} \big) \,
  \Big( 8\,{\Mvariable{i_2}}^3 - 
    12\,{\sqrt{-\Mvariable{i_1} - \Mvariable{i_2}}}\,
     {\sqrt{\Mvariable{i_1} - \Mvariable{i_2}}}\,{\Mvariable{i_2}}^2\,
     \cos (2\,\Mvariable{\varphi_1}) \\ + 
    6\,\Mvariable{i_2}\,\big( -{\Mvariable{i_1}}^2 + {\Mvariable{i_2}}^2
       \big) \,\cos (4\,\Mvariable{\varphi_1}) + 
    {\Mvariable{i_1}}^2\,{\sqrt{-\Mvariable{i_1} - \Mvariable{i_2}}}\,
     {\sqrt{\Mvariable{i_1} - \Mvariable{i_2}}}\,
     \cos (6\,\Mvariable{\varphi_1}) \\ - 
    {\sqrt{-\Mvariable{i_1} - \Mvariable{i_2}}}\,
     {\sqrt{\Mvariable{i_1} - \Mvariable{i_2}}}\,{\Mvariable{i_2}}^2\,
     \cos (6\,\Mvariable{\varphi_1}) \Big)
\end{multline*} 
and 
\begin{multline*}
  f _2 := 2\,\pi \,\Gamma \, \Big( {\Mvariable{i_1}}^2 + 3\,{\Mvariable{i_2}}^2 - 
      4\,{\sqrt{-\Mvariable{i_1} - \Mvariable{i_2}}}\,
       {\sqrt{\Mvariable{i_1} - \Mvariable{i_2}}}\,\Mvariable{i_2}\,
       \cos (2\,\Mvariable{\varphi_1}) \\
       - 2\,\big( {\Mvariable{i_1}}^2 - {\Mvariable{i_2}}^2 \big) \,
       \cos (4\,\Mvariable{\varphi_1}) \Big)^2
\end{multline*}

Notice that $ \bar{ H} _2 $ is again cyclic in $ \varphi _2 $.  (We remark that this is not true for the original $ H _2 $.)  Hence, from the averaging principle, we obtain a family of one-degree-of-freedom systems, parametrized by the constant parameters $ \epsilon, i_2 $, having a two-dimensional phase space in the coordinates $ \{ i _1, \varphi _1 \} $ and governed by the hamiltonian $ \bar{ H} := H _0 (i_1, \varphi _1; \i _2 ) + \epsilon ^2 \bar{ H _2 } (i_1, \varphi _1; i _2) $.  This is interpreted as describing the dynamics, in the averaged approximation, of a 3-vortex problem where one of the vortices is a binary of radius $\epsilon$.

\section{Conclusions}
\label{sec:conclusions}

We have exploited the (degree minus one) homogeneous character of the equations of motion in the square distances of the four-point vortex problem, together with energy and inertia considerations, to obtain various characterizations of candidate evolutions leading to partial or total collision.  Also, for the general $N$-point vortex problem, we showed that \emph{regular} total collisions imply that the virial of the system is zero.  Motivated by the relationship between candidate total and partial collapse we began the study of a four-vortex system containing two vortices close to binary collapse using the method of hamiltonian averaging.  A detailed study of the conditions and time scales under which the averaged approximation remains valid is material of a forthcoming paper.  In that work we plan to also generalize the reduction procedure described in this paper.


\end{document}